\documentclass[preprint]{aastex}
\usepackage[dvips]{color}

\slugcomment{}

\shorttitle{New $BJHK_{\rm s}$-Band Transits and Stellar Variability of GJ~1214 System}
\shortauthors{Narita et al.}

\begin{document}

\title{
Multi-Color Transit Photometry of GJ~1214b through \textit{BJHK}$_{\rm s}$-Bands\\
and a Long-Term Monitoring of the Stellar Variability of GJ~1214
}


\author{Norio Narita\altaffilmark{1,2}, Akihiko Fukui\altaffilmark{3}, Masahiro Ikoma\altaffilmark{4}, 
Yasunori Hori\altaffilmark{1}, Kenji Kurosaki\altaffilmark{4}, Yui Kawashima\altaffilmark{4},
Takahiro Nagayama\altaffilmark{5},
Masahiro Onitsuka\altaffilmark{1,2}, Amnart Sukom\altaffilmark{1,2}, Yasushi Nakajima\altaffilmark{6},
Motohide Tamura\altaffilmark{1,7}, Daisuke Kuroda\altaffilmark{3}, Kenshi Yanagisawa\altaffilmark{3},
Teruyuki Hirano\altaffilmark{8}, Kiyoe Kawauchi\altaffilmark{8},  Masayuki Kuzuhara\altaffilmark{8},
Hiroshi Ohnuki\altaffilmark{8}, Takuya Suenaga\altaffilmark{1,2},  Yasuhiro H. Takahashi\altaffilmark{1,7},
Hideyuki Izumiura\altaffilmark{2,3}, Nobuyuki Kawai\altaffilmark{9}, Michitoshi Yoshida\altaffilmark{10}
}

\altaffiltext{1}{National Astronomical Observatory of Japan, 2-21-1 Osawa, Mitaka, Tokyo 181-8588, Japan}
\altaffiltext{2}{The Graduate University for Advanced Studies, Shonan Village, Hayama, Kanagawa 240-0193, Japan}
\altaffiltext{3}{Okayama Astrophysical Observatory, National Astronomical Observatory of Japan, 
Asakuchi, Okayama 719-0232, Japan}
\altaffiltext{4}{Department of Earth and Planetary Science, The University of Tokyo, 7-3-1 Bunkyo-ku, Tokyo 113-0033, Japan}
\altaffiltext{5}{Department of Astrophysics, Nagoya University, Furo-cho, Chikusa-ku, Nagoya 464-8602, Japan}
\altaffiltext{6}{Hitotsubashi University, 2-1 Naka, Kunitachi, Tokyo 186-8601, Japan}
\altaffiltext{7}{Department of Astronomy, The University of Tokyo, 7-3-1 Hongo, Bunkyo-ku, Tokyo 113-0033, Japan}
\altaffiltext{8}{Department of Earth and Planetary Sciences, Tokyo Institute of Technology, 2-12-1 Ookayama, Meguro-ku, Tokyo 152-8551, Japan}
\altaffiltext{9}{Department of Physics, Tokyo Institute of Technology, 2-12-1, Oookayama, Meguro, Tokyo 152-8551, Japan}
\altaffiltext{10}{Hiroshima Astrophysical Science Center, Hiroshima University 1-3-1, Kagamiyama, Higashi-Hiroshima, Hiroshima 739-8526, Japan}
\email{norio.narita@nao.ac.jp}


\begin{abstract}
We present 5 new transit light curves of GJ~1214b taken in $BJHK_{\rm s}$-bands.
Two transits were observed in $B$-band using the Suprime-Cam and the FOCAS instruments
onboard the Subaru 8.2m telescope, and one transit was done
in $JHK_{\rm s}$-bands simultaneously with the SIRIUS camera
on the IRSF 1.4m telescope.
MCMC analyses show that the planet-to-star radius ratios are,
$R_{\rm p}/R_{\rm s} = 0.11651 \pm 0.00065$ ($B$-band, Subaru/Suprime-Cam),
$R_{\rm p}/R_{\rm s} = 0.11601 \pm 0.00117$ ($B$-band, Subaru/FOCAS),
$R_{\rm p}/R_{\rm s} = 0.11654 \pm 0.00080$ ($J$-band, IRSF/SIRIUS),
$R_{\rm p}/R_{\rm s} = 0.11550 ^{+0.00142}_{-0.00153}$ ($H$-band, IRSF/SIRIUS), and
$R_{\rm p}/R_{\rm s} = 0.11547 \pm 0.00127$ ($K_{\rm s}$-band, IRSF/SIRIUS).
The Subaru Suprime-Cam transit photometry shows a possible spot-crossing feature.
Comparisons of the new transit depths and those from previous studies with the theoretical models
by \citet{2012ApJ...756..176H} suggest that the high molecular weight atmosphere
(e.g., 1\% H$_2$O + 99\% N$_2$) models are most likely, however,
the low molecular weight (hydrogen dominated) atmospheres with extensive clouds are
still not excluded.
We also report a long-term monitoring of the stellar brightness variability of GJ~1214
observed with the MITSuME 50cm telescope in $g'$-, $R_\mathrm{c}$-,
and $I_\mathrm{c}$-bands simultaneously.
The monitoring was conducted for 32 nights spanning 78 nights in 2012,
and we find a periodic brightness variation
with a period of $P_{\rm s} = 44.3\pm1.2$ days and semi-amplitudes of 2.1\%$\pm$0.4\% in $g'$-band,
0.56\%$\pm$0.08\% in $R_\mathrm{c}$-band, and 0.32\%$\pm$0.04\% in $I_\mathrm{c}$-band.
\end{abstract}

\keywords{planetary systems -- planets and satellites: atmosphere --
planets and satellites: individual(GJ1214b) -- stars: individual(GJ1214) -- techniques: photometric}

\section{Introduction}

Super-earths are an emerging population of extrasolar planets whose masses and radii lie
between those of the Earth and the Uranus/Neptune.
The nature of super-Earths, such as internal structure and atmospheric compositions,
remains almost unknown since there is no super-Earth in our Solar System.
Transiting super-Earths are thus invaluable targets for observations to learn
the nature of super-Earths in details.

GJ~1214b discovered by \cite{2009Natur.462..891C} is the first-ever transiting super-Earth
around an M dwarf
which enables us to study its atmosphere through so-called transmission spectroscopy,
thanks to the small host star's size ($\sim0.2R_{\odot}$).
For the purpose, a number of observers have measured transit depths of GJ~1214b
in various wavelength regions (e.g., \citealt{2010Natur.468..669B, 2011ApJ...743...92B,
2011ApJ...736...78C, 2011ApJ...731L..40D, 2011ApJ...730...82C, 2011ApJ...736...12B,
2012ApJ...747...35B, 2012A&A...538A..46D, 2013PASJ...65...27N,
2013ApJ...765..127F, 2013MNRAS.431.1669T}), and have narrowed down possible
atmospheric models proposed by theorists
(e.g., \citealt{2010ApJ...716L..74M, 2012ApJ...756..176H, 2012ApJ...753..100B}).

Among recent publications, \citet{2012ApJ...756..176H} reported various
atmospheric models and compared them with the previous observations.
They concluded that a hydrogen-rich atmosphere with a haze of small ($\sim0.1\mu$m)
particles is the most likely model for GJ~1214b, however, they also noted that
this is only valid if the Rayleigh scattering feature (a rise of transit depths in short optical wavelength)
claimed by \citet{2012A&A...538A..46D} in the optical $g$-band is true.
\citet{2012ApJ...756..176H} also reported that if the short-wavelength result
is inaccurate, then alternative likely models are (1) an N$_2$ and water dominated atmosphere,
(2) a solar-abundance (hydrogen dominated) atmosphere,
with thick clouds at or above the 1 mbar level,
or (3) a solar-abundance atmosphere with a haze of $\ge1\mu$m particles.
In those alternative cases, however, deeper $K_{\rm s}$-band transits claimed by
\citet{2011ApJ...736...78C} and \citet{2012A&A...538A..46D} are incompatible
with the models.
Thus the discussions of likely atmospheric models are largely depend on
the reliability of the results by \citet{2011ApJ...736...78C} and \citet{2012A&A...538A..46D}.

Motivated by this fact, we previously tried simultaneous $JHK_{\rm s}$-transit photometry
of GJ~1214b using the SIRIUS camera on the IRSF 1.4m telescope in 2011
\citep{2013PASJ...65...27N}.
Consequently, we did not find a deeper transit in $K_{\rm s}$-band; namely our
result was inconsistent with the results by \citet{2011ApJ...736...78C} and
\citet{2012A&A...538A..46D}, and instead support a shallower transit
reported by \citet{2011ApJ...743...92B}.
This result has raised the possibility of the alternative models by \citet{2012ApJ...756..176H}.

Moreover, \citet{2013MNRAS.431.1669T} recently presented new $g'$- and $V$-band transits,
which was shallower than the $g$-band transit by \citet{2012A&A...538A..46D}.
Their results were still consistent with \citet{2012A&A...538A..46D},
but also consistent with no Rayleigh scattering (water-dominated atmosphere) model
due to large uncertainty in $R_{\rm p}/R_{\rm s}$.
Although the $g$-band result by \citet{2012A&A...538A..46D} could not be explained
without the Rayleigh scattering,
the new results by \citet{2013MNRAS.431.1669T}
raised fundamental questions whether or not the Rayleigh scattering feature is actually present.
Thus the argument of the presence of the Rayleigh scattering in the atmosphere of
GJ~1214b is still unsolved.

More recently, \citet{2013ApJ...765..127F} presented new Spitzer photometry and also
conducted comparisons of various atmospheric models with all of the previous observations.
They found that the best-fit model was one of the alternative models in \citet{2012ApJ...756..176H},
which contains 1\% H$_2$O + 99\% N$_2$ with a thick tholin haze of $0.1\mu$m particles.
They also mentioned that a pure water model and a flat line (no atmosphere) model
are still acceptable at the time of their publication.

Based on the previous discussions, we consider that remaining important keys to
distinguish atmosphere models of GJ~1214b are
(1) confirmation of a rise of transit depths in optical blue region due to the Rayleigh scattering,
and (2) further confirmation of $K_{\rm s}$-band transit depths compared to $J$-band ones.
To address the above problems, we conducted $B$-band
(bluer than $g$- and $g'$-band) transit observations with the Subaru 8.2m telescope
to confirm or constrain the Rayleigh scattering feature,
and also conducted follow-up transit observations in $JHK_{\rm s}$-bands with
the IRSF 1.4m telescope in South Africa to check the transit depths
in those bands once again.

Meanwhile it is also important to learn and estimate the systematic effects
due to the stellar variability for implications of transit depths.
Previously \citet{2011ApJ...736...12B} reported $\sim3.5$ mmag stellar brightness variability
in MEarth band (similar to $i+z$ band) with a period of $\sim53$ days
and $\sim7$ mmag variability in $V$ band with a period of $\sim41$ days. 
However, no independent confirmation of the stellar variability was reported.
For this reason, we additionally monitored long-term stellar brightness variability
with the MITSuME 50cm telescope at Okayama Astrophysical Observatory (OAO)
in $g'$-, $R_c$-, and $I_c$-bands spanning 78 nights
to examine any systematic effect due to the stellar variability of GJ~1214.
All the listed observations were conducted in 2012.

In this paper, we report results of the above new observations and
present discussions on atmosphere models of GJ~1214b.
The rest of the paper is organized as follows,
We summarize our observations and methods of data reductions in Section~2.
We describe analyses of transit light curves and stellar variability in Section~3.
We present results of our analyses and discuss implications of the results in Section~4.
Finally, we summarize this paper in Section~5.

\section{Observations and Data Reductions}

\subsection{Subaru 8.2m Telescope}

We observed two transits of GJ~1214b with the Subaru 8.2m telescope
on the top of Mauna Kea, Hawaii, USA.
We used the Subaru Prime Focus Camera (Suprime-Cam: \citealt{2002PASJ...54..833M})
on 2012 August 12 UT and
the Faint Object Camera and Spectrograph (FOCAS: \citealt{2002PASJ...54..819K})
on 2012 October 8 UT.
Both transits were observed through the Johnson-Cousins $B$-band filter
($0.440\mu {\rm m} \pm 0.054\mu {\rm m} $).

The Suprime-Cam\footnote{http://www.subarutelescope.org/Observing/Instruments/SCam/index.html}
equips a mosaic of ten fully-depleted-type 2K$\times$4K CCDs manufactured
by Hamamatsu Photonics K.K.,
which covers a 34'$\times$27' field of view (FOV) in total with a pixel scale of 0.20'' pixel$^{-1}$
(each CCD has a FOV of 6.8' $\times$ 13.6').
GJ~1214 was observed with the Suprime-Cam during 06:20--10:10 of 2012 August 12 UT.
The condition was photometric through the observations.
The exposure time was set to 40~s and the dead time
(including the CCD readout time of 18~s and other setup times)
was about 29~s (duty cycle of 58\%).
We took 185 frames in total with the Suprime-Cam.
GJ~1214 was located in the 5th CCD chip, named ``satsuki.''
We defocused the telescope so that stars have doughnut-like
point spread function (PSF) to achieve higher photometric precision.
The typical size of the PSF was $\sim15$ pixels ($\sim$3'') in radius.
Primary data reduction, including bias subtraction and flat fielding,
and aperture photometry was carried out with a customized pipeline by
\citet{2011PASJ...63..287F}, with a constant aperture-size mode
where a same aperture size is applied for all images.

The FOCAS\footnote{http://www.subarutelescope.org/Observing/Instruments/FOCAS/index.html}
is installed at the Cassegrain focus of the Subaru telescope.
The camera has a circular FOV of 6' in diameter, covered by
two fully-depleted-type 2K$\times$4K CCDs by Hamamatsu Photonics K.K.
Each CCD has four readout channels and each channel has 512$\times$4176 active pixels.
The pixel scale is 0.104'' pixel$^{-1}$.
Note that the CCDs of the Suprime-Cam and the FOCAS are different.
We observed GJ~1214 with the FOCAS during 04:46--07:08 of 2012 October 8 UT.
The weather on that night was clear, but the observation started just after dusk and
the first-half of the transit was occurred during twilight.
The exposure time was 40~s and the dead time was about 22~s (duty cycle of 65\%).
We obtained 130 frames in total with the FOCAS.
We again defocused the telescope and the typical size of the PSF
was $\sim17$ pixels ($\sim$1.8'') in radius.
Primary reduction for bias subtraction using overscan region was processed with
a dedicated tool named FOCASRED.
Subsequent procedures, such as flat fielding and aperture photometry, were
conducted with the same customized pipeline \citep{2011PASJ...63..287F}
as the case for the Suprime-Cam.

\subsection{IRSF 1.4m Telescope}

We observed a full transit of GJ~1214b with the Infrared Survey Facility (IRSF)
1.4m telescope located in Sutherland, South Africa.\footnote{The IRSF were constructed
and has been operated by Nagoya University, South African Astronomical Observatory (SAAO)
and National Astronomical Observatory of Japan (NAOJ).}
The transit was observed during 19:26--21:40 of 2012 June 14 UT.
We used Simultaneous Infrared Imager for Unbiased Survey
(SIRIUS: \citealt{2003SPIE.4841..459N}) camera for the observation,
which is the same instrument we used in \citet{2013PASJ...65...27N}.
The SIRIUS camera is equipped with
two dichroic mirrors and three
1K$\times$1K HgCdTe detectors, which
can observe $J$- ($1.250\mu {\rm m} \pm 0.085\mu {\rm m} $),
$H$- ($1.63\mu {\rm m}  \pm 0.15\mu {\rm m} $),
$K_{\rm s}$- ($2.14\mu {\rm m}  \pm 0.16 \mu {\rm m} $)
bands simultaneously.
The FOV of the SIRIUS camera is a square of 7.7' on a side and
a pixel scale is 0.45'' pixel$^{-1}$.
The exposure times were set to 40~s and the dead time of the SIRIUS is about 8~s
(duty cycle of 83\%).
We obtained 163 frames on the night.

During observations, we used a position locking software
introduced in \citet{2013PASJ...65...27N}.
Thanks to this software, positions of GJ~1214's centroid on the three detectors
were kept within an rms of about 2 pixels in both X and Y directions.
We note that stellar images were widely defocused so that
the PSF was spread to $\sim$16 pixels ($\sim$7'') in radius.

Data reduction for the IRSF data is carried out with a dedicated pipeline for the
SIRIUS\footnote{http://irsf-software.appspot.com/yas/nakajima/sirius.html},
including a correction for non-linearity, dark subtraction, and flat fielding.
The non-linearity correction enables us to work up to $\sim$25000 ADU with
$\le$1\% linearity, which is sufficient for the current observations \citep{2013PASJ...65...27N}.
Subsequent aperture photometry was done with the pipeline by \citet{2011PASJ...63..287F}.

\subsection{MITSuME 50cm Telescope}

We monitored brightness of GJ~1214 for 32 nights spanning
from 2012 August 15 to 2012 November 1 (spanning 78 nights in total)
with the MITSuME 50cm telescope located in Okayama Astrophysical Observatory, Okayama, Japan.
The purpose of those observations is not for planetary transits but for stellar variability monitoring.
The MITSuME telescope is equipped with
three 1K$\times$1K CCD cameras, which can obtain
$g'$-, $R_\mathrm{c}$-, and $I_\mathrm{c}$-band images simultaneously
\citep{2005NCimC..28..755K,2010AIPC.1279..466Y}.
The each CCD has the pixel scale of 1$''$.5 pixel$^{-1}$ and the FOV is 26$' \times$ 26$'$. 
We obtained about 10--80 frames per clear night.
The exposure time was 60~s and the dead time was 3 s for all bands.
We slightly defocused the MITSuME telescope so that the target did not saturate in $I_\mathrm{c}$-band.
The PSF extends about 2--4 pixels or 3$''$--6$''$ in radius.
We note that the contamination from objects surrounding the target and reference stars are negligible,
because there is no bright object around the stars based on
the 2MASS All-Sky Catalog of Point Sources \citep{2003tmc..book.....C},
and we have checked that the PSFs of the stars do not show significant contamination.
Data reduction and aperture photometry for the MITSuME data are
carried out with the pipeline by \citet{2011PASJ...63..287F} as with the case for the Suprime-Cam.

\section{Analyses}

\subsection{Transit Light Curves}

First, we create dozens of trial light curves using different aperture sizes ($\Delta r$)
and combinations of comparison stars for each observation.
For the comparison stars, we use such stars that are not saturated, nor variable stars,
and in the same CCD chip with GJ~1214.
On this occasion we convert the time system, which is recorded in the FITS headers
in units of Modified Julian Day (MJD) based on Coordinated Universal Time (UTC),
to Barycentric Julian Day (BJD) based on Barycentric Dynamical Time (TDB)
using the algorithm by \citet{2010PASP..122..935E}.
Note that the time for each datum is assigned as the mid-time of each exposure.
We find all the trial light curves $F_{\rm obs}$ exhibit trends at out-of-transit (OOT) phase.
The trends could be caused by slow variability in the brightness of GJ1214 itself or comparison stars,
changing airmass, position changes of the stars on the detectors,
or high sky background, and so on.
We then check the all trial light curves by eye and eliminate obviously poor-quality ones.

Second, in order to select the most appropriate light curve for each observation and 
its baseline correction model for OOT phase, we adopt the Bayesian Information Criteria
(BIC: Schwarz 1978) for our analyses.
The BIC value is given by $\mathrm{BIC} \equiv \chi^2 + k \ln N$,
where $k$ is the number of free parameters, and $N$ is the number of data points. 
We fit the trial light curves with an analytic transit light curve model and various
baseline models simultaneously.

For the transit light curve model, we employ a customized code \citep{2007PASJ...59..763N}
that use the analytic formula by \citet{2009ApJ...690....1O},
which is equivalent with \citet{2002ApJ...580L.171M} when using the quadratic limb-darkening law.
The quadratic limb-darkening law is expressed as
$I(\mu) = 1 - u_1 (1-\mu) - u_2 (1-\mu)^2$,
where $I$ is the intensity and $\mu$ is the cosine of the angle between
the line of sight and the line from the position of the stellar surface to the stellar center.
For the transit model, we fix the orbital period of GJ~1214b to $P = 1.58040481$ days and
the origin of the transit center to $T_{\rm c,0} = 2454966.525123$ BJD$_{\rm TDB}$,
determined by \citet{2011ApJ...743...92B}.
We note that this assumption is justified by the fact that there is no evidence of
significant transit timing variations (e.g., \citealt{2011ApJ...730...82C, 2013ApJ...765..127F}).
We also fix the orbital inclination $i$ to 88.94$^{\circ}$ and
the orbital distance in units of the stellar radius $a / R_{\rm s}$
to 14.9749, which were determined by \citet{2010Natur.468..669B} and
widely adopted in previous studies
\citep{2011ApJ...743...92B, 2011ApJ...736...78C,
2012A&A...538A..46D, 2012ApJ...747...35B, 2013PASJ...65...27N,
2013ApJ...765..127F, 2013MNRAS.431.1669T}.
This assumption is necessary to directly compare transit depths with previous ones.
Empirical quadratic limb-darkening coefficients for $BJHK_{\rm s}$-bands are adopted
from \citet{2011A&A...529A..75C}
(specifically, $u_{\rm 1,B}=0.6366$,
$u_{\rm 2,B}=0.2737$,
$u_{\rm 1,J}=0.0875$,
$u_{\rm 2,J}=0.4043$,
$u_{\rm 1,H}=0.0756$,
$u_{\rm 2,H}=0.4070$,
$u_{\rm 1,Ks}=0.0475$,
$u_{\rm 2,Ks}=0.3502$),
assuming the stellar effective temperature $T_{\rm eff}=3000$ K
and the log of the stellar surface gravity $\log g=5.0$.
These assumptions on $T_{\rm eff}$ and $\log g$
are the same as previous studies (e.g., \citealt{2011ApJ...736...78C}).
We investigate a possible systematic effect due to those assumption
(especially by the limb-darkening parameters) later in section 4.3.
Adopted stellar and planetary parameters are summarized in table~1.
The free parameter for the transit model is thus the radius ratio of
the planet and the star $R_{\rm p}/R_{\rm s}$ only.

For the baseline model functions $F_{\rm oot}$, we assume the following expression
\citep{2013ApJ...770...95F}:
\begin{eqnarray*}
F_{\rm oot} &=& k_0 \times 10^{-0.4\Delta m_{\rm cor}},\\
\Delta m_\mathrm{cor} &=& \sum k_i X_i,
\end{eqnarray*}
where $k_0$ is the normalization factor,
$F_{\rm oot}$ is the baseline flux, $\{{\bf X}\}$ are observed variables,
and $\{{\bf k}\}$ are coefficients.
For the variables $\{{\bf X}\}$, we test various combinations of $t$, $z$, $dx$, $dy$, and $s$, where  
$z$ is airmass, $t$ is time, $dx$ and $dy$ are the relative centroid positions in $x$ and $y$ directions,
$s$ is sky background counts, respectively.

For each trial light curve
(using various aperture sizes and combinations of comparison stars)
and each combination of variables, we optimize free parameters
using the AMOEBA algorithm \citep{1992nrca.book.....P} and evaluate a BIC value.
We then select a light curve which gives the minimum BIC value for each observation.
After this process, we rescale the photometric errors of the data
so that reduced $\chi^2$ for each observation becomes unity.
We also estimate an effect of time-correlated noise
(so-called red noise: \citealt{2006MNRAS.373..231P}) following the methodology by
\citet{2008ApJ...683.1076W}, and find the effect is not significant for the current datasets.
For this reason, we do not further inflate the errors of the data.

Finally, we fit each selected light curve with the transit light curve model and
the baseline correction function model simultaneously.
This is to include systematic uncertainties due to the baseline model
in the planet-to-star radius ratio $R_{\rm p}/R_{\rm s}$.
Free parameters for the fitting are thus $R_{\rm p}/R_{\rm s}$, $k_0$, and selected $\{{\bf k}\}$.
We present all the free parameters for the selected light curves as well as
the aperture sizes used for aperture photometry in Table~2.

To evaluate uncertainties of free parameters,
we use the Markov Chain Monte Carlo (MCMC) method,
following the analysis in \citet{2013PASJ...65...27N}.
We create 3 different chains of 5,000,000 points,
and trim the first 500,000 points from each chain as burn-in.
We set acceptance ratios of jumping for the chains to about 25\%.
We check the convergence of free parameters by the \citet{Gelman92} test
(the Gelman-Rubin convergence diagnostic is less than 1.05).
We define $1\sigma$ uncertainties by the range of parameters between
15.87\% and 84.13\% of the merged posterior distributions.
The results are described in section 4.1.

\subsection{Stellar Variability Monitoring}

For the MITSuME data, first we eliminate data that were taken in high airmass (over 2)
and during predicted transit times.
We then select one comparison star for each band
that meets following conditions:
(1) not a variable star (confirmed by other comparison stars),
(2) brighter than the target but not saturated ($\sim100$ stars for $g'$ band,
$\sim50$ stars for $R_\mathrm{c}$ band, and $\sim20$ stars for $I_\mathrm{c}$ band), and
(3) gives a light curve with the smallest rms.
The reason why we choose only one comparison star and do not use combinations of stars
is one brighter comparison star is sufficient to achieve a good precision ($\sim1$\%)
for our purpose (namely, any combination of comparison stars do not give
significantly higher precision).
We note that we also confirm that our result presented in the subsequent section
(periodicity and amplitudes of GJ~1214's variability) is robust to several choices of a single
comparison star.
In this process, we also remove outliers that separate beyond 3$\sigma$
from mean brightness of each night.
Consequently we use 2646 data in total.

We model the stellar variability of GJ~1214 with a sine curve
following \citet{2011ApJ...736...12B}.
We assume the following expression for the stellar variability,
\begin{eqnarray*}
F = k_{\rm 0, j} \times 10^{-0.4\, k_{\rm z, j}\, z} + A_{j} \times \sin (2\, \pi \, (t - t_0) / P_{\rm s}),
\end{eqnarray*}
where $k_{\rm 0, j}$ are the normalization factors for each band ($j = g', R_\mathrm{c}, I_\mathrm{c}$),
$k_{\rm z, j}$ are the coefficients for the airmass, $A_{j}$ are the semi-amplitudes of the stellar variability,
$t_0$ is the time of zero phase, and $P_{\rm s}$ is the period of the stellar variability.
The free parameters are $k_{\rm 0, j}$, $k_{\rm z, j}$, $A_{j}$, $t_0$, and $P_{\rm s}$ (11 parameters in total).
We search best-fit parameters giving the lowest $\chi^2$ and
largest $\Delta \chi^2$ compared to a null hypothesis 
by minimizing $\chi^2$ using the AMOEBA algorithm.
Note that we set a prior constraint on the time of zero phase as $6150 < t_0 < 6200$ to
make the fitting convergent.
To estimate uncertainties, after the best-fit parameters are determined,
we rescale the photometric errors of the data
so that reduced $\chi^2$ for the fitting becomes unity.
We note that the error rescaling factors are 1.37, 1.27, 1.13 for
$I_\mathrm{c}$-, $R_\mathrm{c}$-, $g'$-bands, respectively.
The uncertainties are estimated by a criterion of $\Delta \chi^2 = 1.0$.
In addition, we also conduct periodogram analyses to show
that we get a unique period in the observing span.
The fitting and periodogram results are shown in section 4.2.

\section{Results and Discussions}

\subsection{Planet-to-Star Radius Ratios}

Table~2 summarizes the best-fit parameters and their uncertainties for each observation
based on the MCMC analyses.
As a result, we obtain the following planet-to-star radius ratios:
$R_{\rm p}/R_{\rm s} = 0.11651 \pm 0.00065$ ($B$-band, Subaru/Suprime-Cam),
$R_{\rm p}/R_{\rm s} = 0.11601 \pm 0.00117$ ($B$-band, Subaru/FOCAS),
$R_{\rm p}/R_{\rm s} = 0.11654 \pm 0.00080$ ($J$-band, IRSF/SIRIUS),
$R_{\rm p}/R_{\rm s} = 0.11550 ^{+0.00142}_{-0.00153}$ ($H$-band, IRSF/SIRIUS), and
$R_{\rm p}/R_{\rm s} = 0.11547 \pm 0.00127$ ($K_{\rm s}$-band, IRSF/SIRIUS).
The observed light curves with the best-fit models for the Subaru and IRSF data
are plotted in Figure~1 -- 5.
We also plot OOT-normalized light curves in Figure~1 -- 5 for reference,
although we fit the transit model and the baseline model simultaneously.

Overall, our observations indicate a flat transmission spectrum through $BJHK_{\rm s}$ bands.
The transit depths do not appear significantly deeper in $B$- and $K_{\rm s}$-bands than
$J$- or $H$-bands.
The current IRSF/SIRIUS results are consistent with our previous ones
with the same instrument \citep{2013PASJ...65...27N},
again refuting the deeper transit in $K_{\rm s}$-band.
Our two $B$-band observations with Subaru Suprime-Cam and FOCAS are
well consistent each other, however, we should note one thing as follows.
In the residuals of Figure~1 ($B$-band, Subaru/Suprime-Cam),
we see a small bump in the early half of the transit.
The feature could be caused by a spot-crossing event.
If the feature is truly a spot-crossing event,
we can learn what effect the spot would have on Rp/Rs by
eliminating the spot region from the data.
For this purpose, we repeat the MCMC analysis using such data
(specifically, data during 2456151.8194 -- 2456151.8277 BJD$_{\rm TDB}$ are removed).
Consequently we get $R_{\rm p}/R_{\rm s} = 0.11882 \pm 0.00070$ for this case.
Although the feature is well consistent with a crossing over a small spot
(a bump-height of $\sim$0.1\% and a time-scale of $\sim$8 min:
\citealt{2011ApJ...730...82C,2011ApJ...736...12B}),
we cannot refute that it is a product of systematic effects.
In addition, we should consider a possibility for a possible 
``hot-spot'' (plage) occultation.
Such a event may occur in stars with strong spot activity such as GJ~1214
(see e.g., \citealt{2013arXiv1304.2140M}; Colon \& Gaidos in prep.).
For the current case, we removed only a possible spot-crossing region
(with upward residuals), but there is a more subtle dip (downward residuals)
just after the possible spot-crossing feature.
As plages are typically located around dark spots, the slight dip may be
caused by a plage crossing.
If this is true, the above radius ratio ($R_{\rm p}/R_{\rm s} = 0.11882 \pm 0.00070$)
should be considered as an upper limit.
As we cannot decisively diagnose whether those features are real or not,
the result with the Subaru Suprime-Cam should be treated with caution.

\subsection{Stellar Variability}

Figure~6 plots the observed MITSuME data and the best-fit sinusoidal models.
Figure~7 presents periodograms for the MITSuME data.
Table~3 presents the best-fit values and their uncertainties for the free parameters.
We obtain the lowest $\chi^2$ at the period of $P_{\rm s} = 44.3$ days
with an uncertainty of $\pm1.2)$ days.
The periodicity is a unique one (no other significant $\Delta \chi^2$ peak)
in the observing span as we can see in figure~7.
The estimated semi-amplitudes of the brightness variability are 0.32\%$\pm$0.04\%
in $I_\mathrm{c}$-band, 0.56\%$\pm$0.08\% in $R_\mathrm{c}$-band,
and 2.1\%$\pm$0.4\% in $g'$-band.
As we can see, the quality of the $g'$-band data are relatively low (compared to the other bands)
due to the faintness of the target.
We thus note that the semi-amplitude of the $g'$-band variability may be still inaccurate.
Those results are very similar to the previous results by \citet{2011ApJ...736...12B},
who reported 3.5 mmag brightness variability
in $i+z$-band with a period of $\sim53$ days and
7 mmag variability in $V$-band at a period of $\sim41$ days. 

To assess the significance of the variability, we calculate $\chi^2$ and BIC values
for both the best-fit case and the null hypothesis case
($P$ and $t_0$ are removed from the model, and
$A_{I_\mathrm{c}}$, $A_{R_\mathrm{c}}$, and $A_{g'}$ are fixed to zero).
We find $\Delta \chi^2 = 161.4$ and $\Delta$BIC = 122.0 ($\Delta k = 5$ and $N = 2646$).
Thus the brightness variability is significantly detected, and
our MITSuME monitoring independently confirms the stellar variability of GJ~1214.
We caution, however, as \citet{2011ApJ...736...12B} mentioned, that the true rotation
period of GJ~1214 could instead be a positive integer multiple of the quoted period.
Since our monitoring covers only 78 nights, we cannot exclude a possibility of
a longer stellar rotation period.

Even though the true stellar rotation period cannot be determined,
the apparent stellar variability derived by our monitoring
is useful to estimate and constrain systematic effects due to the stellar variability
in the transit depths observed in 2012.
In Figure~6, we show the observing dates of the transits,
i.e., 2012 June 14 for the IRSF SIRIUS, 2012 August 12 for the Subaru Suprime-Cam, and
2012 October 8 for the Subaru FOCAS, with vertical lines.
Since the MITSuME monitoring started after the Subaru Suprime-Cam observation,
the phases for the IRSF SIRIUS observation and the Subaru Suprime-Cam observation
are extrapolated by the period of $P_{\rm s} = 44.3$ days.
According to the figure, the brightness of GJ~1214 is 
nearly peak at the IRSF SIRIUS observation,
middle at the Subaru Suprime-Cam observation,
and bottom at the Subaru FOCAS observation.

\subsection{Possible Impacts of Adopted Assumptions on Radius Ratios}

We have adopted some assumptions in our analyses as shown in table~1.
Since any large systematic effects would affect discussions on atmospheric models
of GJ~1214b, we should check the robustness of our results against the assumptions.
In the previous study, we have already confirmed the robustness of the radius ratios
against $P$, $T_{\rm c,0}$, $i$, $a/R_{\rm s}$, and limb-darkening parameters
in $JHK_{\rm s}$-bands \citep{2013PASJ...65...27N}.
Thus a main concern for this study may arise from the assumption of the effective temperature
and corresponding limb-darkening parameters of GJ~1214 in $B$-band.
Indeed, \citet{2013A&A...551A..48A} recently reported
a new parallax measurement for GJ~1214 that makes it appear
that its effective temperature might be hotter than 3000~K (3252~K$\pm$20~K).
Thus the adopted value of the limb-darkening parameters
might become a source of a systematic effect.

To learn the level of the systematic effect and a possible dependence of radius ratios
on the assumption of the effective temperature,
we repeat the MCMC analyses for the $B$-band datasets for following 3 test cases:
(1) $u_1$ is fixed to the value for $T_{\rm eff}=3000$~K and $u_2$ is free,
(2) limb-darkening parameters with $T_{\rm eff}=3200$~K
(specifically, $u_1 = 0.4749$ and $u_2 = 0.3666$: \citealt{2011A&A...529A..75C}) are assumed, and
(3) limb-darkening parameters with $T_{\rm eff}=2800$~K
(specifically, $u_1 = 0.8687$ and $u_2 = 0.0996$: \citealt{2011A&A...529A..75C}) are assumed.

For the case (1),
we find $u_2 = 0.20\pm0.06$ and $R_{\rm p}/R_{\rm s} = 0.11728 \pm 0.00088$
for the Subaru/Suprime-Cam data, while
$u_2 = 0.30\pm0.06$ and $R_{\rm p}/R_{\rm s} = 0.11564^{+0.00150}_{-0.00157}$
for the Subaru/FOCAS data.
The derived $u_2$ values are almost consistent with the empirical value of
$u_2 = 0.2737$ \citep{2011A&A...529A..75C}, and the derived radius ratios
are also consistent with the values reported in table~2 within 1$\sigma$.
We additionally test an MCMC analysis with letting both $u_1$ and $u_2$ free,
and find
$R_{\rm p}/R_{\rm s} = 0.11720 \pm 0.00085$
for the Subaru/Suprime-Cam data, and
$R_{\rm p}/R_{\rm s} = 0.11548^{+0.00155}_{-0.00164}$
for the Subaru/FOCAS data.
Those values are almost the same with the case (1), showing letting one limb-darkening
parameter free is sufficient for this kind of tests.
Based on this test, we estimate that the possible systematic effect due to the limb-darkening
parameters is smaller than $\Delta (R_{\rm p}/R_{\rm s}) \sim0.001$,
and we conclude that our results for radius ratios in $B$-band are robust.
We also note that the above radius ratio values with free limb-darkening parameters
do not change our conclusion in the section 4.5.

From the cases (2) and (3), we find an interesting trend between the radius ratio
and the assumed effective temperature of GJ~1214.
The derived radius ratios are
$R_{\rm p}/R_{\rm s} = 0.11803 \pm 0.00065$ (case 2, Subaru/Suprime-Cam),
$R_{\rm p}/R_{\rm s} = 0.11838 \pm 0.00123$ (case 2, Subaru/FOCAS),
$R_{\rm p}/R_{\rm s} = 0.11441 \pm 0.00070$ (case 3, Subaru/Suprime-Cam),
$R_{\rm p}/R_{\rm s} = 0.11264 \pm 0.00116$ (case 3, Subaru/FOCAS).
Namely, an assumption of a hotter effective temperature gives a larger radius ratio,
and vice versa.
Among the assumed effective temperatures, the case for $T_{\rm eff}=3000$~K
seems the most consistent with the test case (1), thus we do not change our main results.
Although identifying a reason of this trend is beyond the scope of this paper,
this kind of tests for the systematic effect due to the limb-darkening parameters
would be necessary in future studies, especially in bluer wavelength regions.

\subsection{Possible Impacts of Unocculted Starspots on Radius Ratios}

Unocculted starspots are known to cause a systematic effect on an apparent radius ratio
(see e.g., \citealt{2011ApJ...730...82C, 2011A&A...526A..12D, 2011MNRAS.416.1443S}).
The systematic difference of the radius ratio $\Delta (R_{\rm. p}/R_{\rm s})$
caused by the stellar variability due to starspots can be written as
\begin{equation}
\Delta (R_{\rm p}/R_{\rm s}) \simeq  0.5\,\, \Delta f(\lambda)\,\, (R_{\rm p}/R_{\rm s}),
\end{equation}
where $\Delta f(\lambda)$ is stellar brightness variability at wavelength $\lambda$
\citep{2011MNRAS.416.1443S, 2013PASJ...65...27N}.

The brightness of GJ~1214 in $g'$-band on the two observing nights of the Subaru Suprime-Cam
and the Subaru FOCAS is different by $\sim$2\% based on the MITSuME monitoring, and
given that the semi-amplitude of the stellar variability in $B$-band is similar to that in $g'$-band,
a systematic difference between the Subaru Suprime-Cam and the Subaru FOCAS
observations is $\Delta (R_{\rm p}/R_{\rm s}) \sim0.0012$.
However, this value may be too conservative, since the semi-amplitude of the $g'$-band variability
may be inaccurate due to poor signal-to-noise ratio, as we have cautioned in the previous section, 
and since it appears to be much larger than the semi-amplitude in $V$-band ($\sim7$ mmag)
reported by \citet{2011ApJ...736...12B}.
For this reason, we also try an independent simple estimate for the effect of
unocculted spots in $B$-band as follows.
First, we assume the temperatures of the (normal) stellar surface and the spot region to be
$T_{\rm star}=3000$ K and $T_{\rm spot}=2700$ K.
Assuming the black-body profile for the emission from those regions,
we search for an optimal spot coverage that can explain the semi-amplitude of
$R_\mathrm{c}$- and $I_\mathrm{c}$-band variability.
We find that a spot coverage of 0.73\% gives the best-fit for those two bands.
Using this value, the semi-amplitude of the stellar variability in $B$-band is estimated as 0.88\%
and the possible systematic difference in the radius ratio is
$\Delta (R_{\rm p}/R_{\rm s}) \sim0.00053$ at most.
In either case, we estimate a possible systematic effect on the radius ratio in $B$-band
is as small as or smaller than $\Delta (R_{\rm p}/R_{\rm s}) \sim0.001$.

We note that the more unocculted spots exist, the deeper transits would be observed
\citep{2011ApJ...730...82C}.
Namely the transit at the time of the FOCAS observation is expected to be the deepest.
Our results shown in Table~2, however, appear to be inverse, but this is not significant
when considering the uncertainties in $R_{\rm p}/R_{\rm s}$ for the Suprime-Cam
and the FOCAS observations.

For the near-infrared region, if we suppose that the semi-amplitude of the stellar variability in the
near-infrared region is similar or smaller than the variability in $I_\mathrm{c}$-band from
the MITSuME monitoring,
then the difference of the stellar brightness in $JHK_{\rm s}$-bands is less than $\sim$0.6\%.
Based on this assumption, we estimate maximum systematic differences of radius ratios
in $JHK_{\rm s}$-bands are $\Delta (R_{\rm p}/R_{\rm s}) \le0.0004$,
which is well smaller than the observational uncertainties.
On the other hand, if we again adopt the same estimation method used for $B$-band,
we derive the semi-amplitudes of the stellar variability in $JHK_{\rm s}$-bands as
0.20\%, 0.15\%, and 0.12\%, respectively.
Those values are consistent with the above assumption.
Based on the values, we estimate possible systematic effects on the radius ratio
as $\Delta (R_{\rm p}/R_{\rm s}) \sim0.00012 (J), 0.00009 (H),$ and $0.00007 (K_{\rm s})$,
respectively.
Thus we conclude that we can neglect systematic effects due to unocculted spots
in the near-infrared region within the current observational errors.

\subsection{Atmospheric Models}

Measured radius ratios of GJ~1214b published so far are plotted with respect to wavelength in
Figure~\ref{fig: atmosphere model}, which also shows the five best-fit spectra from the Figure~21
of \citet{2012ApJ...756..176H}.
As described in Section~1, \citet{2013PASJ...65...27N} raised a possibility of a high molecular weight
(high-$\mu$), vapor-rich atmosphere which predicts a flat spectrum, by showing that a $K_{\rm s}$-band transit
was shallower than those from \citet{2011ApJ...736...78C} and \citet{2012A&A...538A..46D}. 
The new $K_s$-band transit depth from the IRSF SIRIUS observation
is consistent with the previous values of ours and \citet{2011ApJ...743...92B}. 
This does not mean, however, that another possibility of a low-$\mu$, hydrogen-dominated
atmosphere is excluded when considering extensive clouds.
This is because both theoretical spectra of the high-$\mu$ and low-$\mu$ atmospheres
yield similar transit depths at the $K_s$-band wavelength, as seen in Figure~\ref{fig: atmosphere model}a.

The $B$-band transit that we have observed with the Subaru FOCAS is significantly shallower than
the $g$-band transit reported by \citet{2012A&A...538A..46D}.
As discussed in \citet{2012ApJ...756..176H}, the deep $g$-band transit needs
a Rayleigh-scattering-like feature (the Rayleigh slope) in the visible to near-infrared wavelength region,
which appears in the theoretical spectrum of a hydrogen-rich atmosphere with tholin haze 
(see Figure~\ref{fig: atmosphere model}a). 
In contrast, the shallow $B$-band transit is consistent with the model spectra for high-$\mu$ atmospheres
without the Rayleigh slope (see Figure~\ref{fig: atmosphere model}b). 
Among the five best-fit models proposed by \citet{2012ApJ...756..176H},
the model spectra for the 1\% H$_2$O + 99\% N$_2$ atmosphere (with and without tholin haze)
appear most likely.
While the solar-composition (low-$\mu$) atmosphere with opaque clouds may also account for
the FOCAS $B$-band transit as well as the IRSF $JHK_{\rm s}$-band transits.
In contrast, the low-$\mu$ atmosphere with 0.1$\mu$m tholin haze is inconsistent with
the FOCAS $B$-band transit, and the similar atmosphere with 1$\mu$m tholin haze
is also inconsistent with the IRSF $HK_{\rm s}$-band transits. 

Unfortunately, the result from the Suprime-Cam is inconclusive due to the possible spot-crossing event,
as discussed in the section 4.1. 
Given the spot-crossing is real, the difference of the transit depths between the two observations
($R_{\rm p}/R_{\rm s} {\rm (Suprime-Cam)} = 0.11882 \pm 0.00070$ and
$R_{\rm p}/R_{\rm s} {\rm (FOCAS)} = 0.11601 \pm 0.00117$) is about 2$\sigma$
(1$\sigma$ here is a square-root of sum of squares of both uncertainties).
Although the significance of the difference is marginal, the two transit depths appears to be inconsistent.
One possibility to explain the difference is that the stellar activity might cause a temporal change
in the amount of haze in the atmosphere.
In the case of Titan's atmosphere, it is known that the solar UV flux and Saturn's magnetospheric
electrons and protons contribute to synthesize tholin haze \citep{1984Icar...60..127K}. 
When a close-in planet like GJ~1214b passes through a stellar-spot magnetosphere,
the production rate of tholin haze might be affected.
A time variation in the amount of tholin haze might lead to the change of the $B$-band transit depths.
In this case, some amount of a low-$\mu$ component should be present in the atmosphere of GJ~1214b.
Although it is highly speculative, this possibility is worth exploring by repeated future observations.

Whether the atmosphere contains a significant amount of water or not
affects our understanding of the origin of GJ~1214b. 
The deep photometric transit originally presented by \citet{2009Natur.462..891C}
suggests that GJ~1214b contains some low-$\mu$ components. 
If the planet is completely differentiated, it must be enveloped with a low-$\mu$ atmosphere. 
Within the context of the core-accretion model, a hydrogen-rich atmosphere of nebular origin
is the most-likely possibility \citep{2012ApJ...753...66I}. 
Detailed modeling of the internal structure of GJ~1214b \citep{2011ApJ...733....2N, 2013arXiv1305.2629V}
predicts that the hydrogen-rich atmosphere constitutes several \% of the planet mass to
reproduce the mass-radius relationship for this planet.
It should be also noted that recently \citet{2013arXiv1305.4124M} suggested that
cloud and hydrocarbon haze formation in the atmosphere of GJ~1214b by
(non-)equilibrium chemistry favored atmospheric models with enhanced metallicity.

The atmosphere of GJ~1214b may have been subject to photo-evaporative
mass loss due to stellar XUV irradiation. 
Although the current irradiation level would be so low that the atmosphere will undergo no significant
mass loss today and in the future, several studies advocated that a GJ~1214b-like planet had experienced
a significant removal of the atmosphere in the past (e.g. \citealt{2013arXiv1303.3899O}). 
The mass loss history is, however, not well-constrained, because we do not know the current
intrinsic luminosity of GJ 1214b, which affects the speed of the thermal evolution significantly. 
Thus, we cannot deny the presence of such a hydrogen-rich atmosphere theoretically 
from the viewpoint of the internal structure and evolution.

On the other hand, absence of low-$\mu$ components in the atmosphere would
give some important constraints on the structure and origin of this planet. 
To reconcile with the planet's low-density, one possibility would be that low-$\mu$
components should be incorporated in the interior;
for example, the envelope in the mixed state with water and H/He.
Giant collisions between super-Earths or heavy secondary bombardments of
volatile-rich planetesimals would be needed for such mixing to occur.
Such processes triggered in a planetary system would lead to form multiple planets.
Provided that this picture holds true for GJ~1214b, we predict that additional planets should
exist in the GJ~1214 system.
Thus search for outer planets helps us to learn the formation history of GJ~1214b.

\subsection{Suggestions for Future Observations}

Although our results have suggested that GJ~1214b has
a fairly flat transmission spectrum through $BJHK_{\rm s}$-bands,
it is still difficult to determine one decisive atmosphere model.
Experiences have shown that broadband single-color transit photometry is not efficient
to constrain an atmosphere model in the presence of starspots and the stellar variability.
More effective ways to characterize atmospheres of transiting planets would be
(1) simultaneous multi-band transit photometry using small-medium ground-based telescopes
(e.g., \citealt{2011ApJ...736...78C,2012A&A...538A..46D,2013PASJ...65...27N,2013ApJ...770...95F}),
(2) multi-object spectro-photometry using large ground-based telescopes
(e.g., \citealt{2010Natur.468..669B,2011ApJ...743...92B}), and
(3) spectro-photometry using space telescopes
(e.g., \citealt{2012ApJ...747...35B}).
It would be important to observe the wavelength region where the difference of transit depths
between the low-$\mu$ and the high-$\mu$ atmospheres is significant,
especially the Rayleigh slope (optical) region and around $K$-band region.
In addition, repeated transit observations are highly desirable to improve the significance and
to check possible time variations.
As the ongoing ground-based transit surveys (e.g., MEarth) and
the future space-based survey like TESS \citep{2010AAS...21545006R}
will discover more transiting super-Earths around nearby cool host stars,
the current experiences for GJ~1214b would become a good practice for the future.

\section{Summary}

We have presented two $B$-band transits observed with the Suprime-Cam and FOCAS on
the Subaru 8.2m telescope
and one simultaneous $JHK_{\rm s}$-band transit taken with the SIRIUS camera on
the IRSF 1.4m telescope.
Our measurements of transit depths suggest a fairly flat transmission spectrum
through $BJHK_{\rm s}$-bands.
Comparisons of our new results and previous observations with theoretical atmospheric models
from \citet{2012ApJ...756..176H} indicate that the high-$\mu$ (water-rich) atmosphere models are
most likely, although the low-$\mu$ (hydrogen-dominated) atmosphere with thick clouds may
still account for the observations.
As noted in section 4.1., our Subaru Suprime-Cam data show a possible spot-crossing event.
Suppose that the spot-crossing is real and the data from the event is removed,
our two $B$-band results are marginally inconsistent.
It is slightly puzzling, but it may be simply due to unknown systematic effects, or it may be explained
by temporal changes of transit depths due to the stellar activity and thereby time variations
of haze amount.
To further constrain the atmosphere model of GJ~1214b and to check a possibility of
the presence of time variations in transit depths or the Rayleigh slope,
more repeated observations described in the previous subsection (section 4.6.) would be desirable in the future.

\acknowledgments

This paper is partly based on data collected at Subaru Telescope,
which is operated by the National Astronomical Observatory of Japan (NAOJ).
We acknowledge kind supports by Fumiaki Nakata for the Subaru Suprime-Cam observation,
and by Takashi Hattori for the Subaru FOCAS observation.
The IRSF project was financially supported by the Sumitomo foundation
and Grants-in-Aid for Scientific Research on Priority Areas (A) (Nos.
10147207 and 10147214) from the Ministry of Education, Culture,
Sports, Science and Technology (MEXT) of Japan.
The operation of IRSF is supported by Joint Development Research of
National Astronomical Observatory of Japan, and  Optical Near-Infrared
Astronomy Inter-University Cooperation Program, funded by the MEXT.
We are also grateful to Kouji Ohta and Shogo Nagayama for operations of the MITSuME telescope.
We thank Eric Gaidos and Knicole Colon for fruitful discussions on GJ~1214b.
N.N. acknowledges supports by NAOJ Fellowship,
NINS Program for Cross-Disciplinary Study, and
Grant-in-Aid for Scientific Research (A) (No.~25247026) from the MEXT.
M.I. is supported by Grant-in-Aid for Scientific Research (C) (No.~25400224) from the MEXT
and by Challenging Research Award from Tokyo Institute of Technology.
This work is in part supported by Grant-in-Aid for JSPS Fellows
(Nos. 23-3491, 23-271, 25-3183, 25-8826) from the MEXT.
K.K. is supported by a grant for the Global COE Program,
``From the Earth to Earths'', by the MEXT.
M.T. is supported by the MEXT, Grants-in-Aid No. 22000005.
H.I. is supported by JSPS KAKENHI No. 23244038.
N.K. is supported by Grant-in-Aid for Creative Scientific Research No. 14GS0211 from the MEXT.

{\it Facilities:} \facility{Subaru}. \facility{IRSF}. \facility{MITSuME}.

%%%%%%%%%%%%%%%%%%%%%%%%%%%%%%%%%%%%%%%%%%%%%%%%%%%%%%%%%%%%%%%%%%%%%%

%%%%%%%%%%%%%%%%%%%%%%%%%%%%%%%%%%%%%%%%%%%%%%%%%%%%%%%%%%%%%%%%%%%%%%

%%%%%%%%%%%%%
\begin{deluxetable}{lcc}
\tabletypesize{\small}
\tablecaption{Assumed parameters and their sources.}
\tablewidth{0pt}
\tablehead{
\colhead{Parameter} & \colhead{Value} & \colhead{Source} 
}
\startdata
$P$ [days] & 1.58040481  & \citet{2011ApJ...743...92B} \\[2pt]
$T_{\rm c,0}$ [BJD$_{\rm TDB}$] & 2454966.525123 & \citet{2011ApJ...743...92B} \\[2pt]
$i$ [$^{\circ}$] & 88.94 & \citet{2010Natur.468..669B} \\[2pt]
$a/R_{\rm s}$ & 14.9749 & \citet{2010Natur.468..669B} \\[2pt]
$u_{\rm 1,B}$ & 0.6366 & Claret \& Bloeman (2011) \\[2pt]
$u_{\rm 2,B}$ & 0.2737 & Claret \& Bloeman (2011) \\[2pt]
$u_{\rm 1,J}$ & 0.0875 & Claret \& Bloeman (2011) \\[2pt]
$u_{\rm 2,J}$ & 0.4043 & Claret \& Bloeman (2011) \\[2pt]
$u_{\rm 1,H}$ & 0.0756 & Claret \& Bloeman (2011) \\[2pt]
$u_{\rm 2,H}$ & 0.4070 & Claret \& Bloeman (2011) \\[2pt]
$u_{\rm 1,Ks}$ & 0.0475 & Claret \& Bloeman (2011) \\[2pt]
$u_{\rm 2,Ks}$ & 0.3502 & Claret \& Bloeman (2011) \\[2pt]
$T_{\rm eff}$ [K] & 3000 & \citet{2011ApJ...736...78C} \\[2pt]
$\log g$ & 5.0 &  \citet{2011ApJ...736...78C} \\[2pt]
\enddata
\end{deluxetable}
%%%%%%%%%%%%%

\clearpage
%%%%%%%%%%%%%
\begin{deluxetable}{lcc}
\tabletypesize{\scriptsize}
\tablewidth{0pt}
\tablecaption{Best-fit values and uncertainties for the parameters of the transit light curves
based on the MCMC analyses.\label{tabletransit}}
\tablehead{
\colhead{Parameter} & \colhead{Value} & \colhead{Uncertainty} 
}
\startdata
\multicolumn{3}{c}{Subaru/Suprime-Cam (2012 August 12)}\\[1pt]
$R_{\rm p}/R_{\rm s}$ (B) & 0.11651 & $\pm0.00065$\\[1pt]
$k_0$    & 0.98676   &   $\pm0.00099$  \\ [1pt]
$k_t$    &  -0.0110  &  $\pm0.0018$   \\[1pt]
$k_z$    &  0.00494  &  $\pm0.00036$   \\[1pt]
$\Delta r$  & 17  &  -- \\[1pt]
\tableline
\multicolumn{3}{c}{Subaru/Suprime-Cam (spot-feature removed)}\\[1pt]
$R_{\rm p}/R_{\rm s}$ (B) & 0.11882 & $\pm0.00070$\\[1pt]
$k_0$    & 0.98712   &   $\pm0.00092$  \\ [1pt]
$k_t$    &   -0.0104 &  $\pm0.0016$   \\[1pt]
$k_z$    & 0.00482   &  $\pm0.00033$   \\[1pt]
$\Delta r$  & 17  &  -- \\[1pt]
\tableline
\multicolumn{3}{c}{Subaru/FOCAS (2012 October 8)}\\[1pt]
$R_{\rm p}/R_{\rm s}$ (B) & 0.11601 & $\pm0.00117$\\[1pt]
$k_0$    &  0.9780  &  $\pm0.0011$   \\[1pt]
$k_t$    &  -0.0373  &  $\pm0.0075$   \\[1pt]
$k_z$    &   0.00642 &  $\pm0.00039$   \\[1pt]
$k_s$ ($\times10^{-7}$)   &  -1.45  &  $\pm0.37$   \\[1pt]
$\Delta r$  & 18  &  -- \\[1pt]
\tableline
\multicolumn{3}{c}{IRSF/SIRIUS (2012 June 14)}\\[1pt]
$R_{\rm p}/R_{\rm s}$ (J) & 0.11654 & $\pm0.00080$\\[1pt]
$k_0$    &  1.00119  &   $\pm0.00012$  \\[1pt]
$k_t$    &  0.0128  &   $\pm0.0015$  \\[1pt]
$\Delta r$  & 17  &  -- \\[1pt]
\tableline
\multicolumn{3}{c}{IRSF/SIRIUS (2012 June 14)}\\[1pt]
$R_{\rm p}/R_{\rm s}$ (H) & 0.11550 & $^{+0.00142}_{-0.00153}$\\[1pt]
$k_0$    &  0.9835  &  $\pm0.0074$   \\[1pt]
$k_t$    &  0.025  &  $\pm0.012$   \\[1pt]
$k_z$    &  0.0052  &  $\pm0.0022$   \\[1pt]
$k_x$ ($\times10^{-4}$)   &  1.86  & $\pm0,49$    \\[1pt]
$\Delta r$  & 17  &  -- \\[1pt]
\tableline
\multicolumn{3}{c}{IRSF/SIRIUS (2012 June 14)}\\[1pt]
$R_{\rm p}/R_{\rm s}$ (K$_{\rm s}$) & 0.11547 & $\pm0.00127$\\[1pt]
$k_0$    &  0.9856  &   $\pm0.0014$  \\[1pt]
$k_z$    &  0.00495  &  $\pm0.00042$   \\[1pt]
$\Delta r$  & 16  &  -- \\[1pt]
\enddata
\end{deluxetable}
%%%%%%%%%%%%%
%%%%%%%%%%%%%
\begin{deluxetable}{lcc}
\tabletypesize{\small}
\tablecaption{Best-fit values and uncertainties for the parameters of the Sinusoidal Variability of GJ1214.}
%\tablewidth{6cm}
\tablewidth{0pt}
\tablehead{
\colhead{Parameter} & \colhead{Value} & \colhead{Uncertainty} 
}
\startdata
$P_s$ [days] & 44.3  &$\pm 1.2$ \\[2pt]
$t_0$ [JD-2,450,000] & 6174.22 &$^{+0.77}_{-0.81}$\\[2pt]
$A_{I_\mathrm{c}}$ & 0.00319 &$\pm 0.00038$\\[2pt]
$A_{R_\mathrm{c}}$ & 0.00558  &$\pm 0.00078$\\[2pt]
$A_{g'}$ & 0.0213 &$\pm 0.0042$\\[2pt] 
$k_{0, I_\mathrm{c}}$ & 0.9958& $ \pm 0.0015$\\[2pt]
$k_{0, R_\mathrm{c}}$ & 0.9729 & $ \pm 0.0031$\\[2pt]
$k_{0, g'}$ & 1.0405 &$^{+0.0158}_{-0.0166} $\\[2pt]
$k_{z, I_\mathrm{c}}$ & -0.0028 &$ \pm 0.0011$\\[2pt]
$k_{z, R_\mathrm{c}}$ & -0.0210 & $ \pm 0.0023$\\[2pt]
$k_{z, g'}$ & 0.0262 &$^{+0.0127}_{-0.0119} $\\[2pt]
\enddata
\end{deluxetable}
%%%%%%%%%%%%%

\clearpage

%%%%%%%%%%%%%
\begin{figure*}
\epsscale{.9}
\plotone{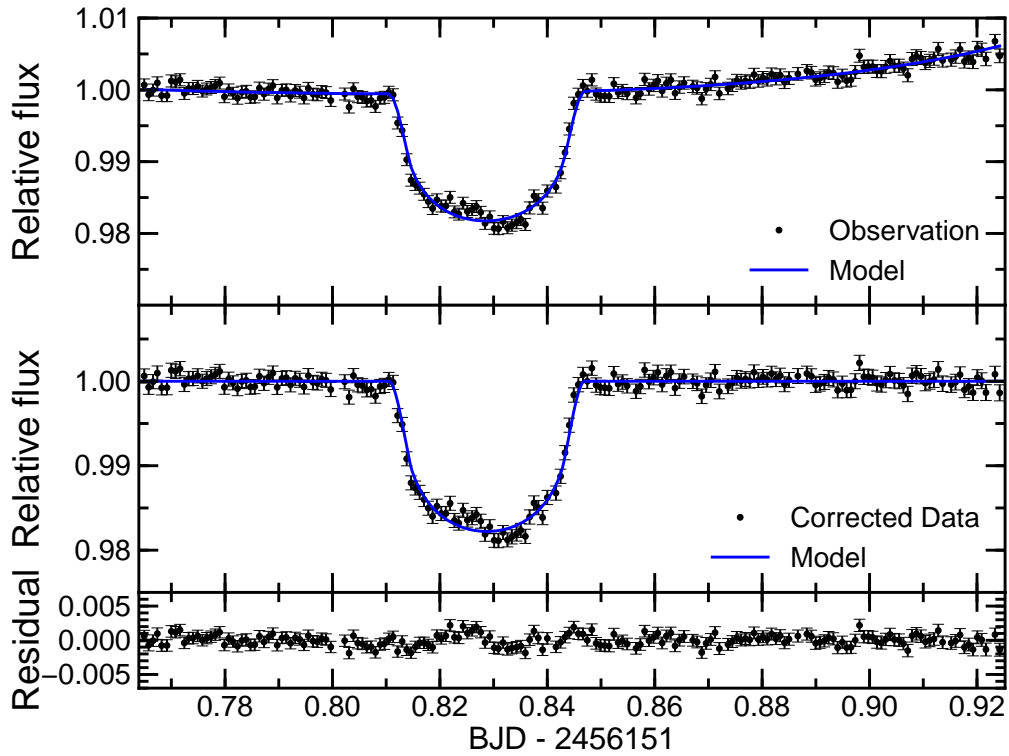}
\caption{Top: a raw fractional light curve and the best-fit model light curve of GJ~1214
taken on 2012 August 12 UT with the Subaru Suprime-Cam in $B$-band.
The model light curve includes both an analytic transit light curve and a baseline function.
Middel: an OOT-normalized light curve and the best-fit transit light curve model.
Bottom: residuals between the observed data and the best-fit model.
}
\end{figure*}
%%%%%%%%%%%%%
\begin{figure*}
\epsscale{.7}
\plotone{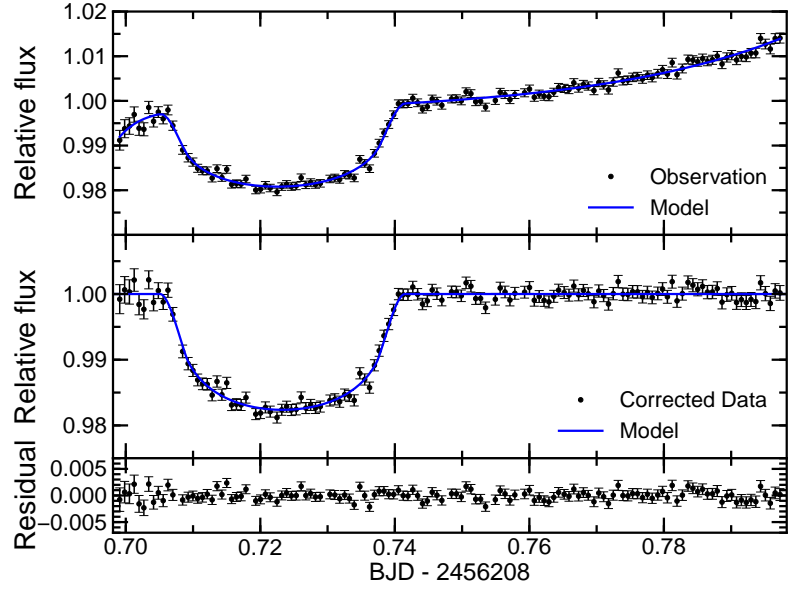}
\caption{Same as figure~1, but for the data taken on 2012 October 8 UT, with the Subaru FOCAS
in $B$-band.}
\end{figure*}
%%%%%%%%%%%%%
\begin{figure*}
\epsscale{.7}
\plotone{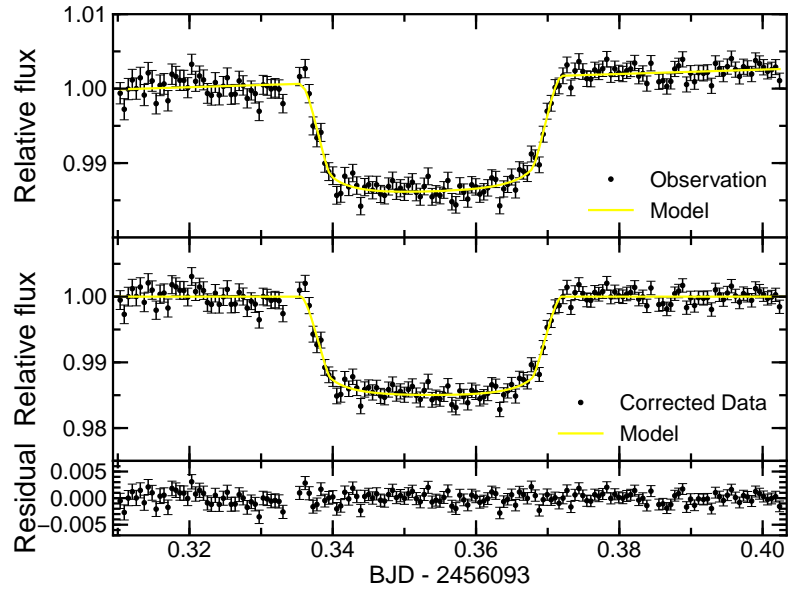}
\caption{Same as figure~1, but for the data taken on 2012 June 14 UT, with the IRSF SIRIUS
in $J$-band.}
\end{figure*}
%%%%%%%%%%%%%
\begin{figure*}
\epsscale{.7}
\plotone{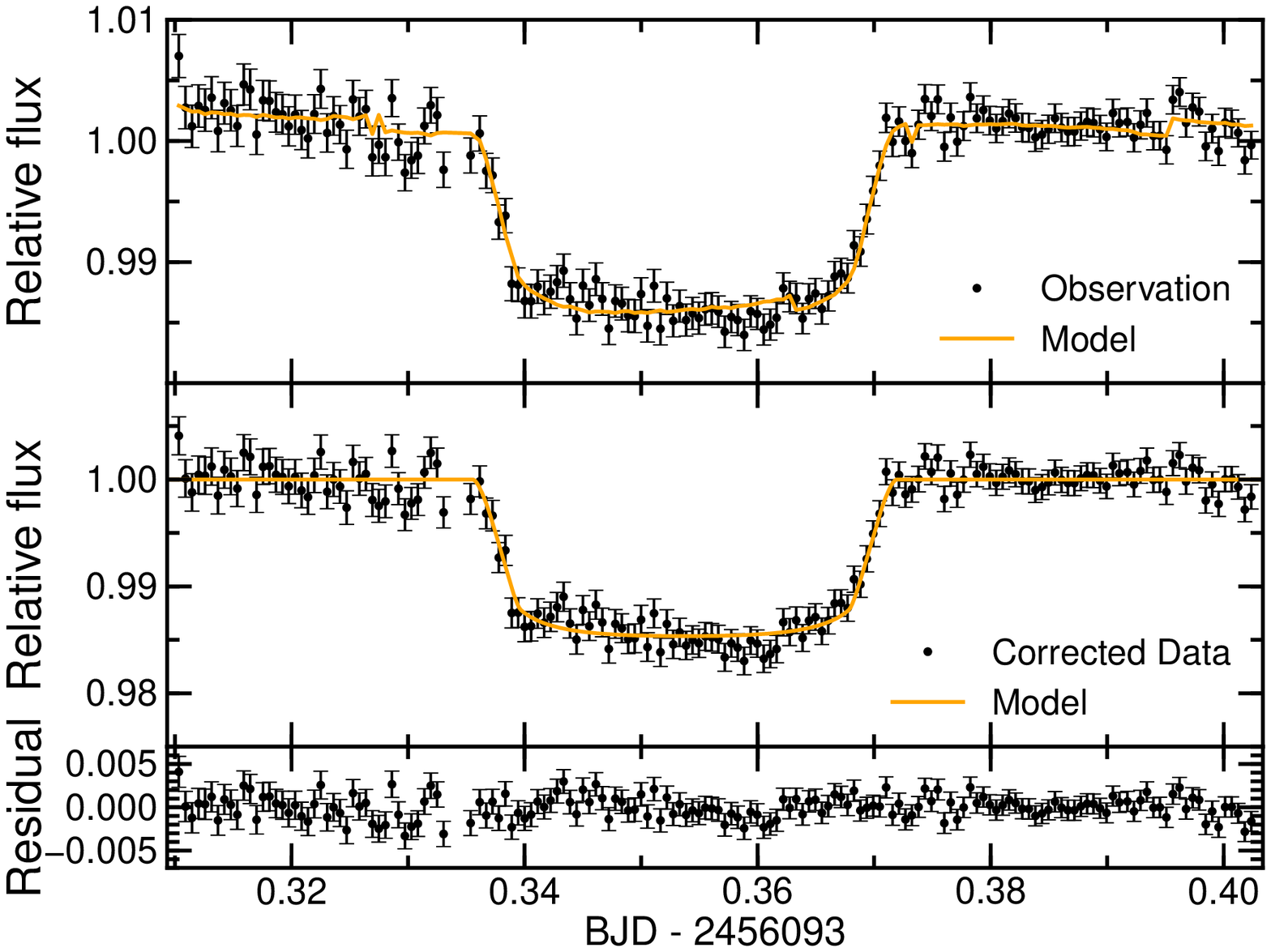}
\caption{Same as figure~3, but for the data taken in $H$-band.}
\end{figure*}
%%%%%%%%%%%%%
\begin{figure*}
\epsscale{.7}
\plotone{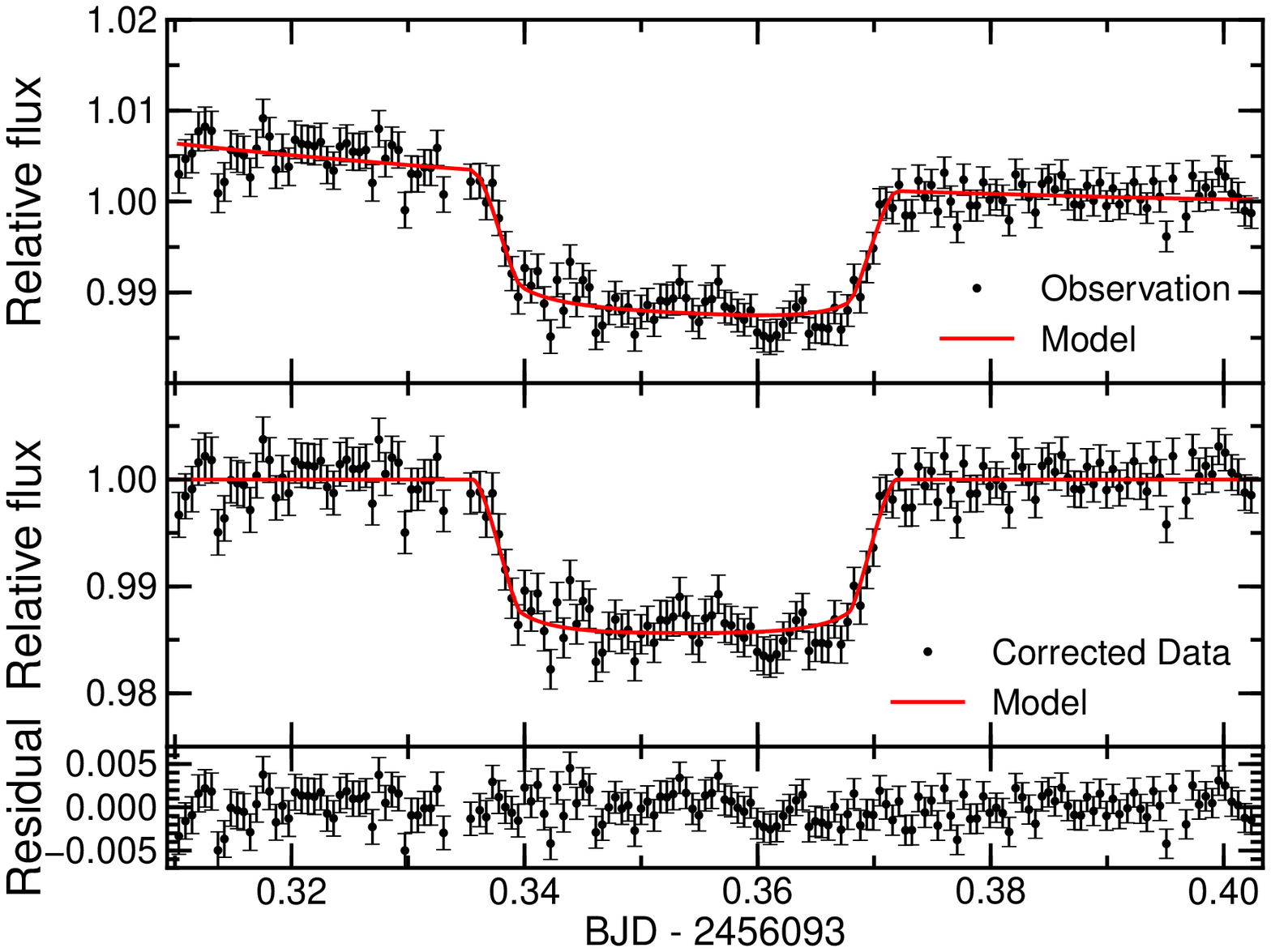}
\caption{Same as figure~3, but for the data taken in $K_{\rm s}$-band.}
\end{figure*}
%%%%%%%%%%%%%
\begin{figure}
\epsscale{.7}
\plotone{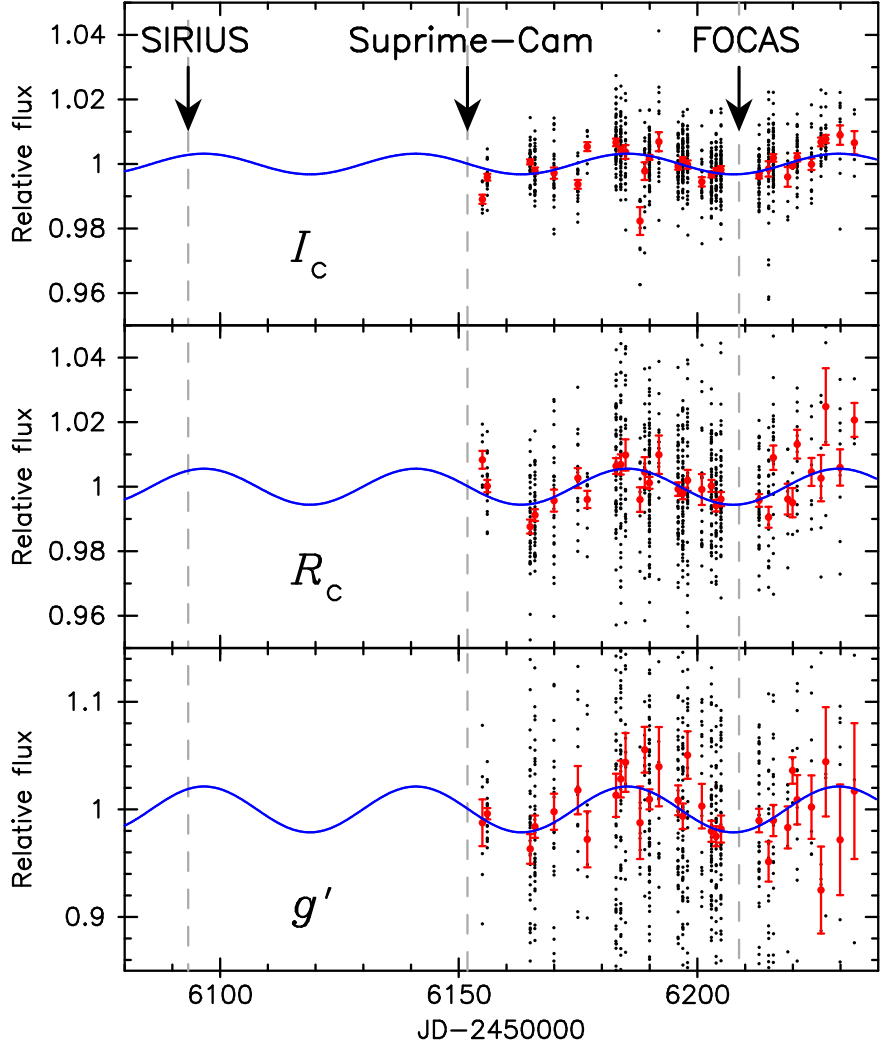}
\caption{Long-term light curves of GJ~1214 obtained with the MITSuME 50cm telescope
at the Okayama Astrophysical Observatory in 2012.
The data were taken in three ($I_\mathrm{c}$: top, $R_\mathrm{c}$: middle, $g'$: bottom)
bands simultaneously.
Observed data are plotted as dots.
Data with error bars indicate mean values and rms divided by $\sqrt{N}$ values,
where $N$ is the number of data points, of the observed data for each night
for reference.
Sinusoidal curves are the best-fit models based on the AMOEBA analysis described in section 3.2.
}
\end{figure}
%%%%%%%%%%%%%
\begin{figure}
\epsscale{.7}
\plotone{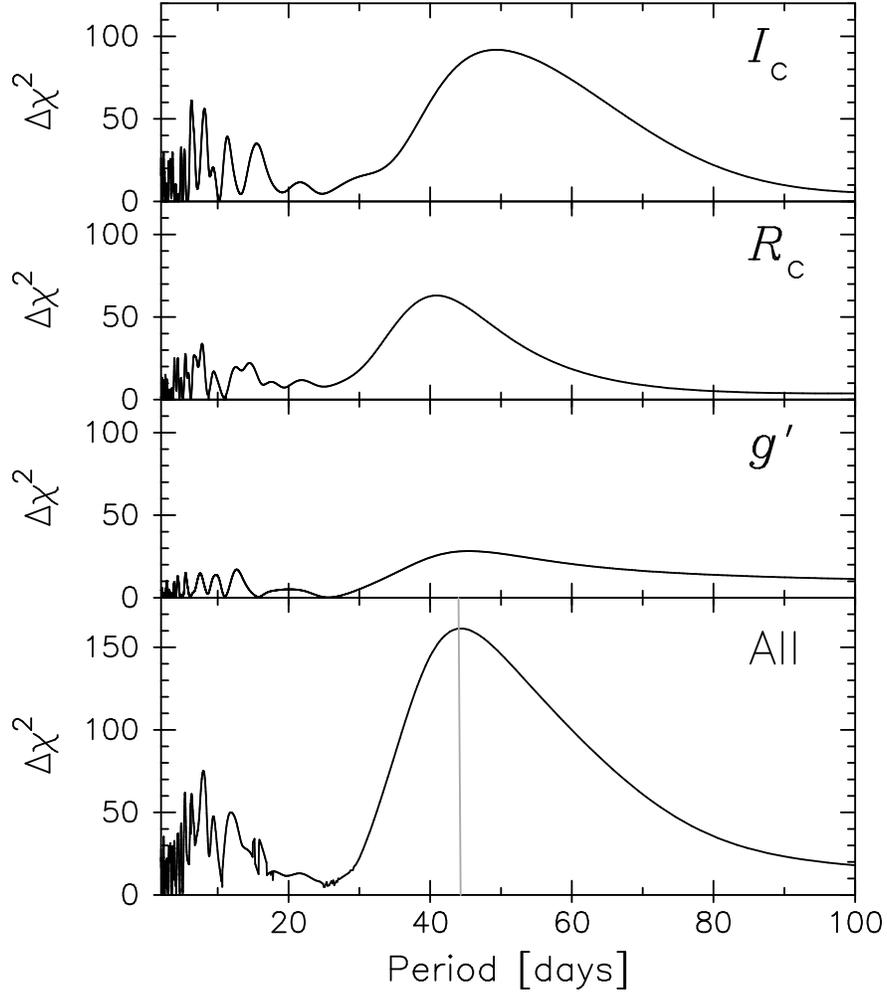}
\caption{Periodograms for the MITSuME data.
Results for $I_\mathrm{c}$-, $R_\mathrm{c}$-, $g'$-bands,
and a combined case are shown from top to bottom.
The vertical axis indicates $\Delta \chi^2$
between the models for no variation ($A_{j} = 0$) and
for a fixed period at the horizontal axis (other parameters are free).
The optimal case ($P_{\rm s} = 44.3$ days) is indicated by the vertical line in the bottom panel.
}
\end{figure}
%%%%%%%%%%%%%

%%%%%%%%%%%%%%
\begin{figure}[b]
  \begin{center}
  \includegraphics[width=14cm,keepaspectratio]{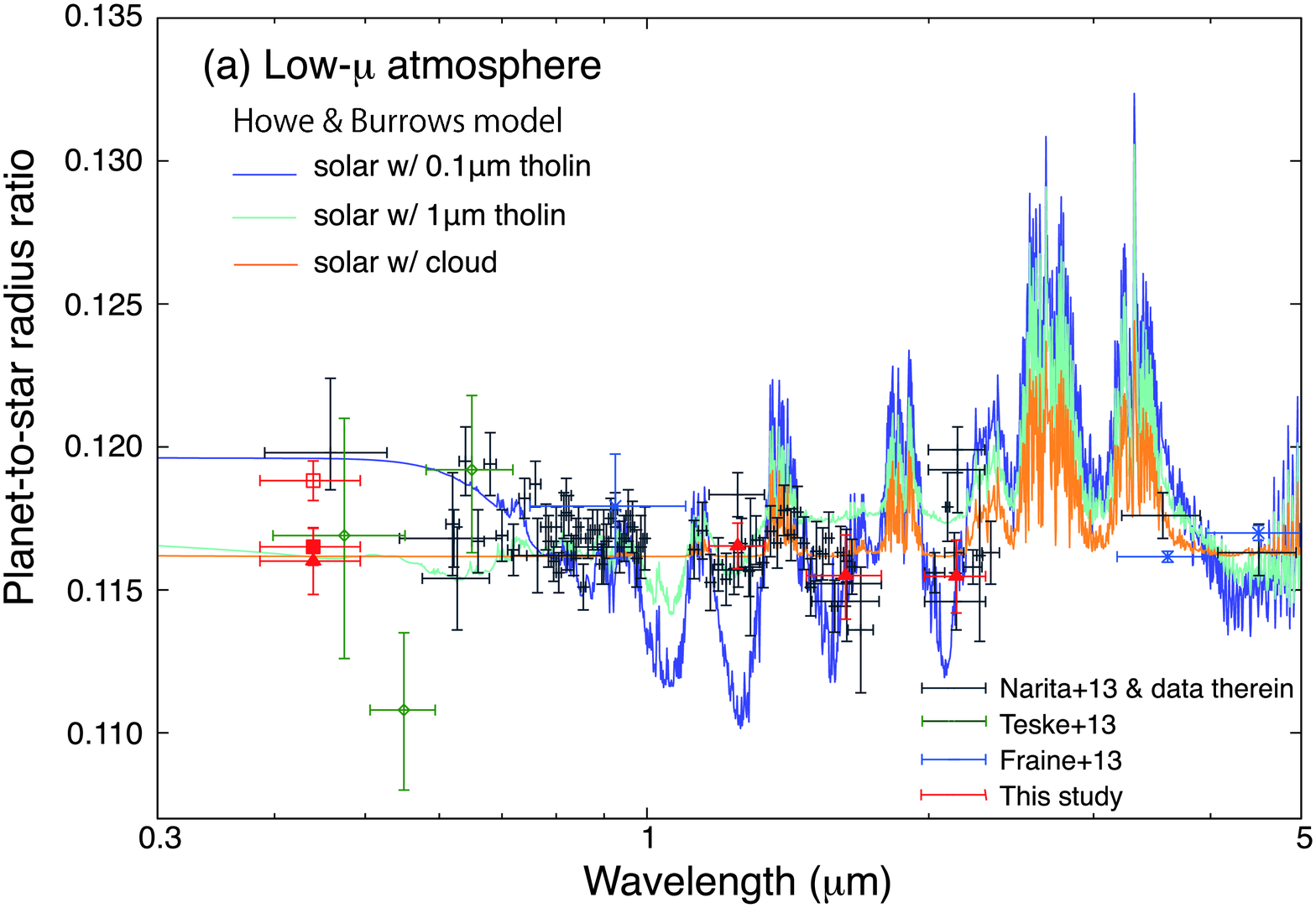}
  \includegraphics[width=14cm,keepaspectratio]{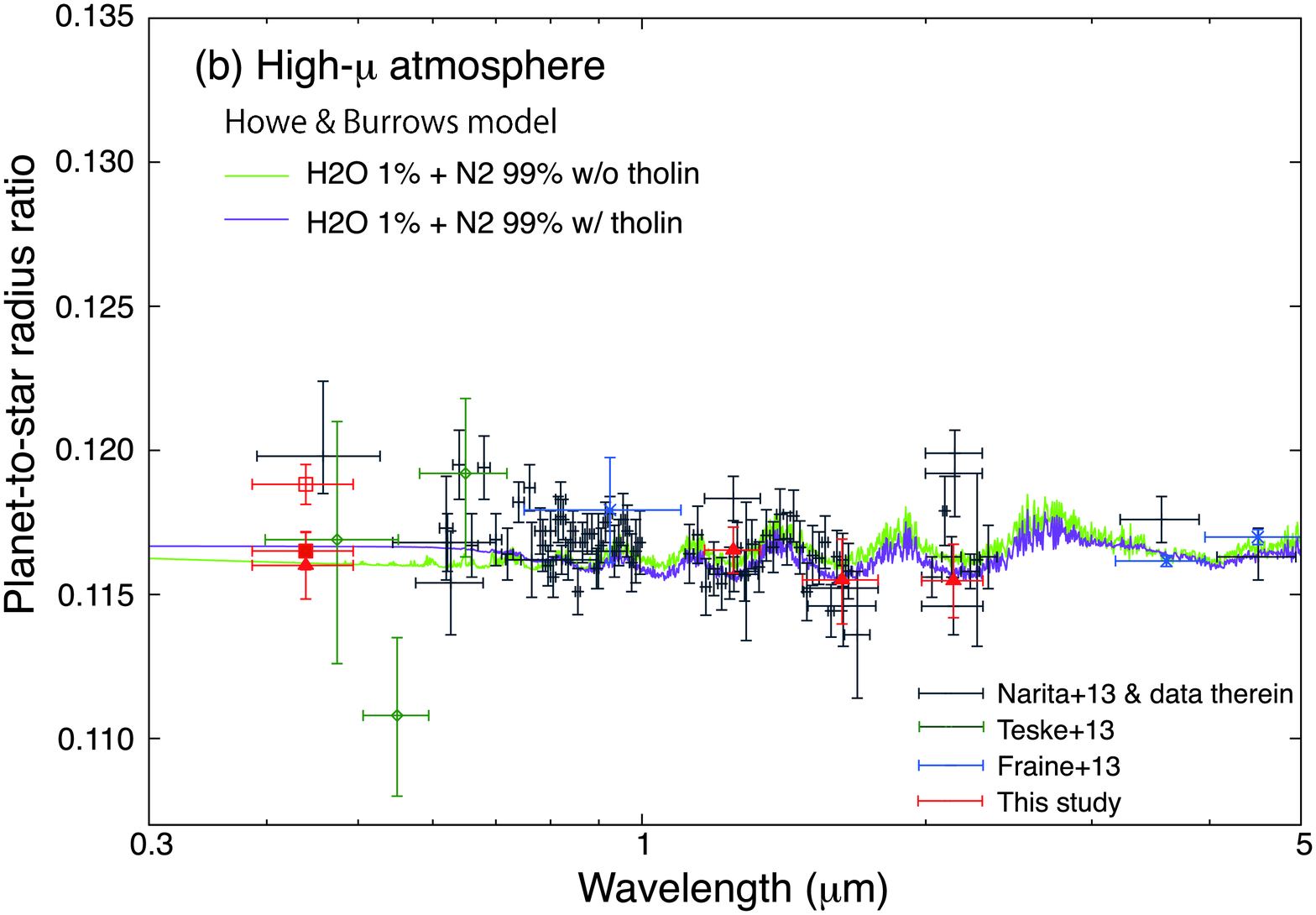}
  \caption{Measured planet-to-star radius ratios vs. wavelength in microns for GJ~1214b, compared with the five best-fit theoretical spectra from Howe \& Burrows~(2012). The same set of observation data is shown in panel (a) and (b). In (a), the theoretical spectra for the solar-abundance atmosphere are shown, while those for the water-rich atmosphere are shown in (b). 
  As for the $B$-band, the filled triangle and square represent the data obtained with the FOCAS and the Suprime-Cam, while the open square does the case for the Suprime-Cam without possible spot-crossing data.}
  \label{fig: atmosphere model}
  \end{center}
\end{figure}
%%%%%%%%%%%%%%

\end{document}